\documentstyle[aps,twocolumn,epsfig]{revtex}
\begin{document}
\newcommand{\ve}[1]{\mbox{\boldmath $#1$}}
\noindent
{\bf Comment on ``Fragmented Condensate Ground State of Trapped Weakly
Interacting Bosons in Two Dimensions"}
\vskip0.5pc 
Recently Liu {\it et al.} \cite{Hui} examined the lowest state of a 
weakly-interacting Bose-Einstein condensate. In addition to other interesting 
results, using the method of the pair
correlation function \cite{Ezra}, they questioned the validity of the
mean-field picture of the formation of vortices and stated that the vortices
are generated at the center of the cloud. This is in apparent contradiction to
the Gross-Pitaevskii approach, which predicts that the vortices successively
enter the cloud from its outer parts as $L/N$ (where $N$ is the number of atoms
in the trap and $\hbar L$ is the angular momentum of the system) increases
\cite{Rokhsar,KMP}. We have managed to reproduce the results of 
Ref.\,\cite{Hui}, however a more careful analysis presented below confirms the
validity of the mean-field approach.

Figures 1 and 2 show equidensity lines of 
\begin{eqnarray}
    P({\bf r},{\bf r}_A) =  \frac
   {\langle \Psi_{L,N} | \sum_{i \neq j}
  \delta({\bf r}-{\bf r}_i) \delta({\bf r}_A-{\bf r}_j)
   | \Psi_{L,N} \rangle}
 {(N-1) \langle \Psi_{L,N} | \sum_j \delta({\bf r}_A-{\bf r}_j)
| \Psi_{L,N} \rangle},
\label{cpd1}
\end{eqnarray}
calculated within the space of $N$ atoms and $L$ oscillator quanta 
and with the additional truncation that only considers single-particle 
components with no radial nodes and angular-momentum quantum numbers $m=0,
\dots, 6$ (as was also assumed in Ref.\,\cite{Hui}.) Here $|\Psi_{L,N} \rangle$
denotes the many-body wavefunction. In these figures ${\bf r}_A=(x,y)=(3,0)$
in units of the oscillator length. The blue colour denotes regions
of lower density, while the red colour denotes regions of higher density.

More specifically, in Fig.\,1 we have taken $N=40$ particles
and $L=28, 32, 36$, and 40, i.e., $L/N=0.7, 0.8, 0.9$, and 1.0.
Also Fig.\,2 shows $P({\bf r},{\bf r}_A)$ for $L=62, 64, 72$,
and 84, i.e., for $L/N=1.55, 1.6, 1.8$, and 2.1.
These figures exhibit two significant differences in comparison with the
conclusions of Ref.\,\cite{Hui}: (i) their qualitative features 
are {\it not} independent of the radial position of the reference
point ${\bf r}_A$, and in fact the important broken symmetry 
observed for $L/N = 0.7, 0.8$, and 0.9 is not visible, unless
$r_A$ is given a significantly large value, and (ii) the entry of 
the first vortex in the region $L/N < 1$, as well as that of the
second vortex in the region $1 < L/N < 1.7$ is clearly seen to be
from the outside and gradually moving inwards with increasing $L/N$.

Finally, Fig.\,1 with $L/N=1$ corresponds
to a single-vortex-like state and is slightly asymmetric with respect 
to the center because of the finite number of atoms we have considered
(the same is also true for the slight asymmetries in the two-fold and 
the three-fold symmetric states shown in Fig.\,2). We have confirmed
numerically that this asymmetry decreases with increasing $N$.

In summary, we disagree with the conclusions of Ref.\,\cite{Hui} in the 
comparison of exact and mean-field description of the mode of entry of 
vortices into the condensate with increasing angular momentum. This does not
however affect the conclusions of Ref.\,\cite{Hui} concerning the fragmentation 
of a rotating condensate.
\\
\noindent 
G. M. Kavoulakis and S. M. Reimann \\
Mathematical Physics, \\
Lund Institute of Technology, P.O. Box 118, \\
S-22100 Lund, Sweden \\ \\
B. Mottelson \\
NORDITA, \\
Blegdamsvej 17, DK-2100 Copenhagen \O, Denmark

\noindent

\begin{figure}
\begin{center}
\epsfig{file=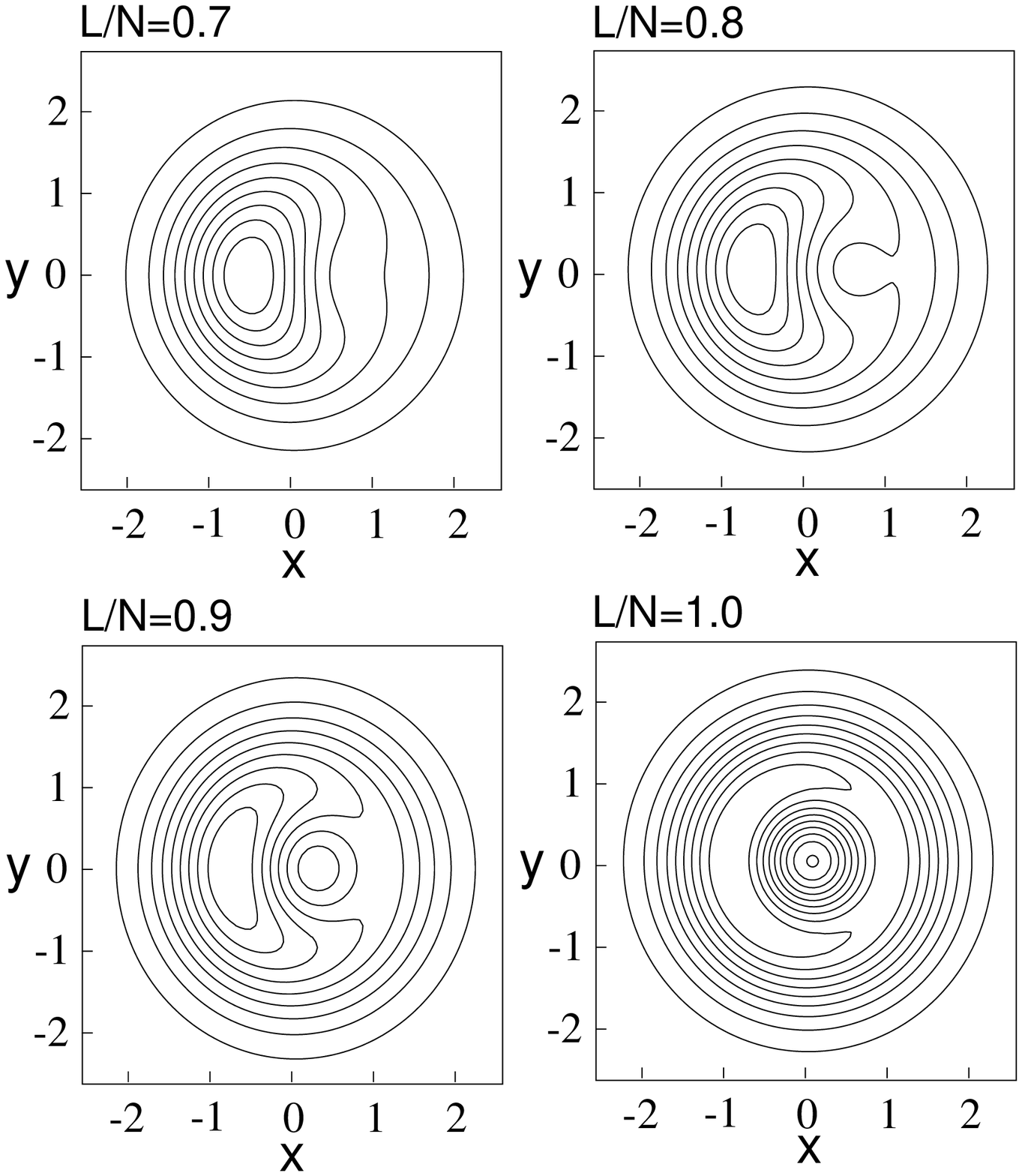,width=5.5cm,height=5.5cm,angle=0}
\vskip 0.5pc
\begin{caption}
{Equidensity lines of $P({\bf r},{\bf r}_A)$ for $N = 40$ and $L=28, 32,
36$, and 40, calculated with the reference point located at ${\bf r}_A=(3,0)$.}
\end{caption}
\end{center}
\label{FIG1}
\end{figure}

\begin{figure}
\begin{center}
\epsfig{file=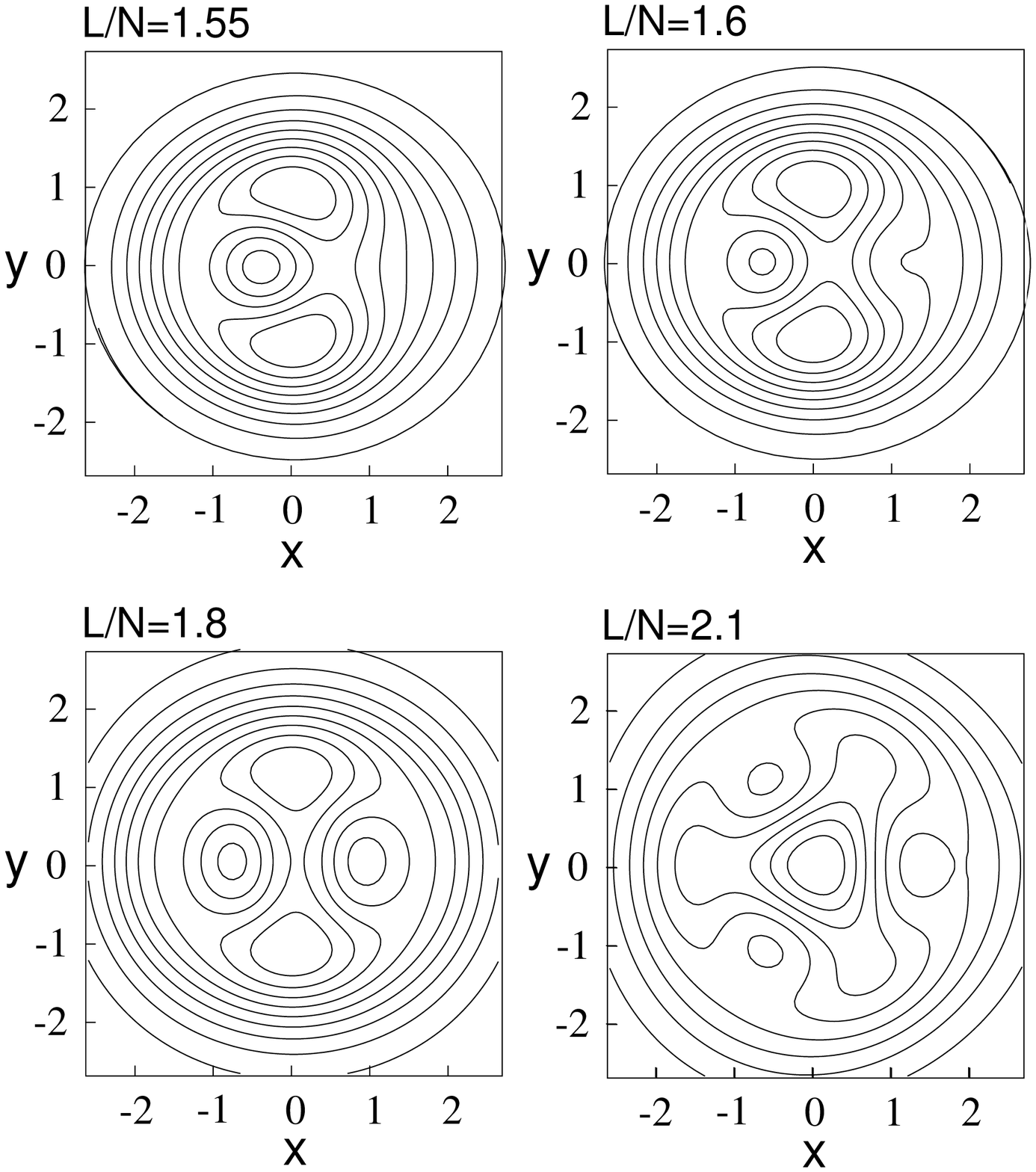,width=5.5cm,height=5.5cm,angle=0}
\vskip 0.5pc
\begin{caption}
{Same as Fig.\,1 for $N = 40$ and $L=62, 64, 72$, and 84.} 
\end{caption}
\end{center}
\label{FIG2}
\end{figure}

\end{document}